\documentclass[%
 reprint,
 amsmath,amssymb,
]{revtex4-1}

\usepackage{graphicx}
\usepackage{dcolumn}
\usepackage{bm}
\usepackage{xcolor}
\usepackage{ulem}

\usepackage{algorithm2e}

\begin{document}

\preprint{APS/123-QED}
\title{Magnetic reconnection as an energy cascade process}

\author{S. Adhikari$^1$}
\email{subash@udel.edu}
 
 \author{T.N. Parashar$^{2,1}$}
\author{M.A. Shay$^{1,3}$}%
\author{W.H. Matthaeus$^{1,3}$} 
\author{P. S. Pyakurel$^4$} 
\author{S. Fordin$^1$} 
\author{J.E. Stawarz$^5$} 
\author{J.P. Eastwood$^5$} 
\affiliation{%
 $^1$Department of Physics and Astronomy, University of Delaware, Newark, Delaware 19716, USA
}%
\affiliation{
$^2$ School of Chemical and Physical Sciences,Victoria University of Wellington, Wellington 6140, NZ
}%
\affiliation{
$^3$ Bartol Research Institute, Department of Physics and Astronomy, University of Delaware, Newark, Delaware 19716, USA
}%
\affiliation{
$^4$ Space Sciences Laboratory, University of California, Berkeley, CA 94720, USA
}%
\affiliation{
$^5$ Department of Physics, Imperial College London, SW7 2AZ, UK
}%

\date{\today}

\begin{abstract}
Reconnection and turbulence are two of the most commonly observed dynamical processes in plasmas, but their relationship is still not fully understood. Using 2.5D kinetic particle-in-cell simulations of both strong turbulence and reconnection, we compare the cross-scale transfer of energy in the two systems by analyzing the generalization of the von K\'arm\'an Howarth equations for Hall magnetohydrodynamics, a formulation that subsumes the third-order law for steady cascade rates. Even though the large scale features are quite different, the finding is that the decomposition of the energy transfer is structurally very similar in the two cases. In the reconnection case, the time evolution of the energy transfer also exhibits a correlation with the reconnection rate. These results provide explicit evidence that reconnection itself is fundamentally an energy cascade process. 
\end{abstract}

\maketitle


\textit{Introduction:} Many naturally occurring and man-made plasmas are observed to be in a turbulent state \cite{coleman1968turbulence, matthaeus1982measurement, bruno2005solar, banerjee2014turbulence, HadidEA18,shi2020dynamical} driven at large scales, either externally or by an energy reservoir.
A nonlinear cascade 
transfers energy from large 
scales to smaller kinetic scales. 
Magnetic reconnection \cite{servidio2009magnetic,yamada2010magnetic},
frequently observed in these systems, is itself 
a nonlinear process, though largely studied independently of 
turbulence. In many cases, turbulence is 
either a consequence or driver of the reconnection process
\cite{matthaeus1986turbulent,servidio2009magnetic, StraussApJ1988, JabbariMNRAS2016}.

Study of the interplay of turbulence and reconnection has 
emerged as a subject of interest. 
For example, for  
reconnection in the presence of turbulence, computed
in two-dimensional ($2$D) incompressible magnetohydrodynamics (MHD) \cite{matthaeus1986turbulent}, 
multiple smaller-scale islands develop, 
the reconnection rate is enhanced, and the energy is isotropized in the spectral space. 
In other
approaches, 
reconnection is 
studied as a subsidiary process occurring in turbulence,
either by diagnosing reconnection occurring in turbulence simulations~\cite{servidio2009magnetic, wan2010accuracy, JabbariMNRAS2016, haggerty2017exploring, papini2019can} or studying how reconnection modifies cascade properties at small scales \cite{EastwoodPRL2009, LoureiroPRL2017, BoldyrevApJ2017, FranciApJL2017, KowalApJ2017, MalletJPP2017,VechApJL2018,ergun2018magnetic}. 
Turbulent features are also studied 
as consequences of instabilities associated with 
large scale reconnection~\cite{EastwoodPRL2009,leonardis2013identification, PucciApJ2018, munoz2018kinetic, LapentaApJ2020}. 

Recent kinetic particle-in-cell (PIC) simulations show that even laminar 2D reconnection exhibits a Kolmogorov $-5/3$ slope in the magnetic spectra, raising the intriguing question: ``Is reconnection itself a cascade, or is it  an independent process often associated with a turbulent cascade?'' In this letter, we analyze the cross-scale energy transfer associated with the von K\'arm\'an Howarth equations generalized to Hall-MHD~\cite{PolitanoPRE1998,hellinger2018karman}, a formulation within which is embedded the famous steady state third-order cascade law~\cite{PolitanoPRE1998} including the Hall effect~\cite{hellinger2018karman,Andres2018energy,BanerjeePRE2020}. We carry out the analysis employing kinetic PIC simulations of both strong turbulence and reconnection, in each case initialized in a regime 
expected to be close to incompressibility. Details of energy transfer in 
both simulations are found to be structurally very similar, giving evidence that reconnection is itself an energy cascade process. Consistent with this conclusion, we also find that the time evolution of the cascade is strongly correlated with the reconnection rate.

\textit{Simulations:}
To study the cascade process in magnetic reconnection, we use two
fully-kinetic 2.5D PIC simulations - a strong turbulence case (Simulation A) and a
laminar reconnection case (Simulation B); see 
Table~\ref{tab:table:sim-info}. 
Time is normalized to the inverse ion cyclotron frequency
($w_{ci}=(eB_0/m_ic)$), where $B_0$ is the normalizing magnetic field.
Length is normalized 
to the ion inertial length $d_{i} = \sqrt{c^2 m_i/(4\pi n_0 e^2)},$ 
where $n_0$ is the normalizing number density.
Speed is therefore normalized to the ion Alfv\'en speed
($v_{A}= d_{i}\,\omega_{ci}$) and temperature to $T_0=m_i v_{A}^2 $. Following standard turbulence notation~\cite{matthaeus1982measurement, biskamp2003magnetohydrodynamic, bruno2013solar} magnetic field ($\mathbf{b} \equiv \mathbf{B}/\sqrt{4\pi m_i n_0}$) and current ($\mathbf{j} \equiv \mathbf{J}/ne$) are normalized to $v_A$.

\begin{table*}[t]
\fontsize{9}{12}\selectfont
\caption{\label{tab:table:sim-info}
Simulation details: Guide field ($B_g$), temperature ($T$), mass ($m$), ions/electrons ($i/e$), the root mean square value ($rms$) $\delta b_{rms} = \sqrt{\langle\lvert \textbf{b} -\langle\textbf{b}\rangle_r \rvert ^2\rangle_r}$, and the fluctuation (turbulence) amplitude $\delta Z = \sqrt{(\delta b_{rms})^2+(\delta u_{rms})^2}$.}
\begin{ruledtabular}
\begin{tabular}{cccccccccccc}
Run  & Type & $L_{box} [d_i]$ & grids & $\Delta x [d_i]$ & $B_g$ & $T_i/T_e$ & $m_i/m_e$ & $n_b$ & $\delta b_{rms}$ & $\delta u_{rms}$ & $\delta Z$  \\
\hline
$A$ & Turbulence & $149.6$ & $4096^2$ & $0.036$ & $1$ & $0.3/0.3$ & $25$ & $1$ & $1/\sqrt{10}$ & $1/\sqrt{10}$ & $1/\sqrt{5}$  \\
\hline
$B$ & Reconnection & $91.59$ & $4096^2$ & $0.022$  & $0$ & $0.05/0.01$ & $25$ & $1$ & $1/\sqrt{5}$ & $0$ & $1/\sqrt{5}$  \\
\end{tabular}
\end{ruledtabular}
\end{table*}

The turbulence simulation is undriven and initially populated with Fourier modes
$k\in [2,4]\times \frac{2\pi}{149.6}$ with average wave number $k_{av} \approx 3\,\frac{2\pi}{149.6} = 0.126$;  
see \cite{parashar2018dependence} for details.
The reconnection simulation is initialized with a double Harris current sheet configuration ( with $k_{av} \approx 2\pi/91.59 = 0.068$ ).
Reconnection is triggered by a 
small magnetic perturbation; see
\cite{adhikari2020reconnection} for details. 
To facilitate the comparison, 
simulation B normalization values are modified from \cite{adhikari2020reconnection} as $B_0 \rightarrow \frac{1}{\sqrt{5}}\,B_0$ and $n_0 \rightarrow 5\,n_0$. With this modification 
simulations A and B have the same background density $n_b$ and the same initial fluctuation amplitude $\delta Z$, as defined in Table~\ref{tab:table:sim-info}.

\begin{figure}[t]
\includegraphics[scale=0.4]{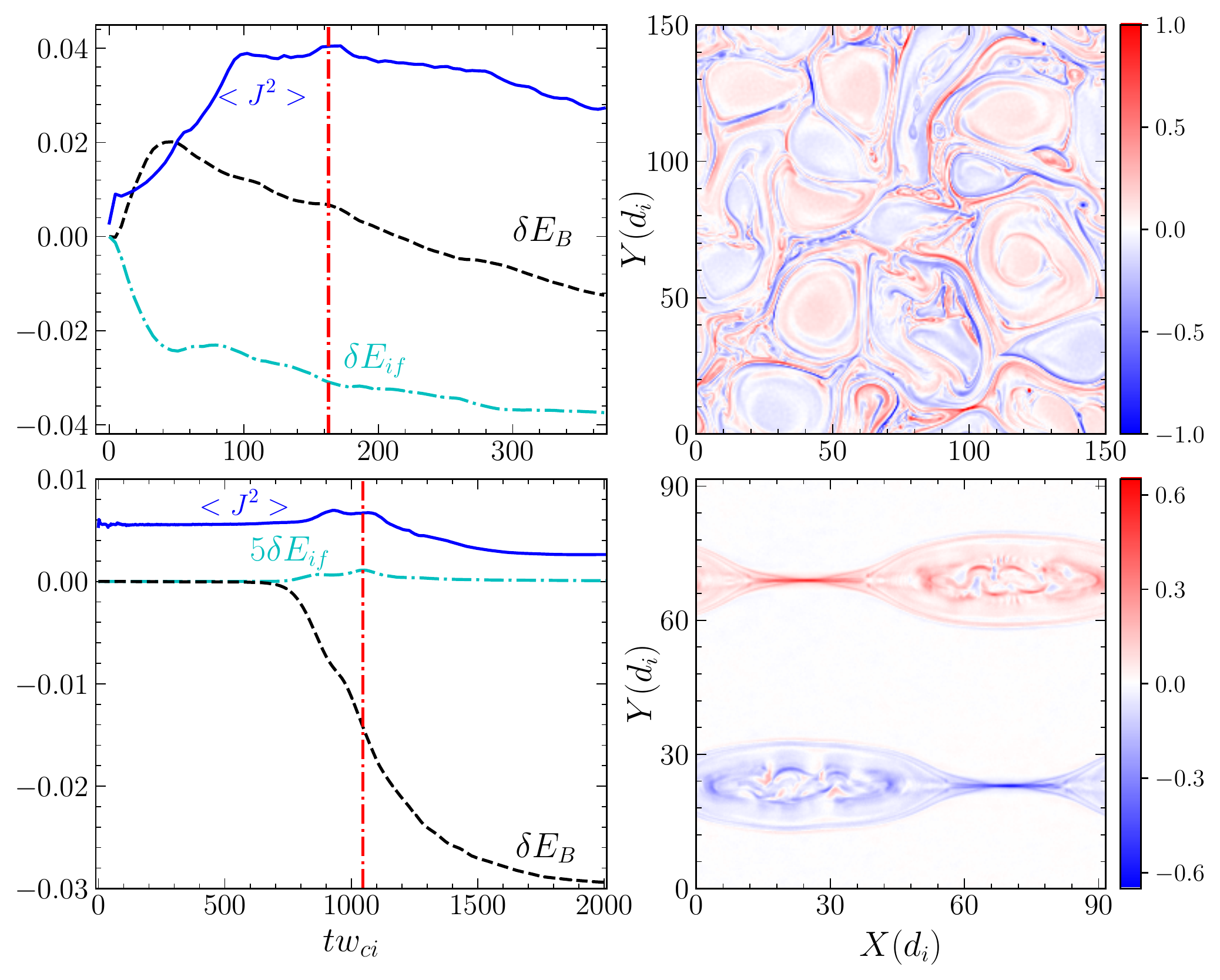}
\caption{\label{fig:sim-overview} Left panel: Evolution of the mean square current (blue), change in magnetic ($\delta E_B$) (black), and ion-flow ($\delta E_{if}$) energy per unit mass (cyan) for simulations A (top) and B (bottom); red vertical line denotes: (sim A) maximum rms $J$ at $t \omega_{ci} = 163$ or (sim B) late quasi-steady reconnection at $t \omega_{ci}=1045.6$. Right panel: $J_z$ of simulation A (top) and B (bottom).}
\end{figure}

Fig.~\ref{fig:sim-overview} provides an overview. 
In the turbulence simulation A, 
an initial Alfv\'enic exchange of energy occurs 
between the ion-flow ($if$) and magnetic field ($B$). 
The fluctuation energy ($E_{if}\,+\,E_B$) decreases monotonically, 
as electrons and ions are heated (see~\cite{parashar2018dependence}).
For simulation B, 
$E_B$ decreases, as  
reconnection transfers most of the energy to thermal energy and a small fraction to $E_{if}$. The mean square current peaks during the quasi-steady phase of reconnection and falls as 
$\delta E_B$ decreases. 
The energy decay rate becomes significant after 
reconnection onset, and peaks during the late quasi-steady phase ($tw_{ci}=1045.6$). 
As evidenced by this figure, the energetics and the currents have a very different evolution in the two cases. \textit{A priori} it appears unlikely that 
the reconnection and turbulence simulations should exhibit similar energy cascades. 

\textit{Cascade Rate:}
To quantify the cascade rate in the simulations, we employ a
particular form of the Hall-MHD von K\'arm\'an Howarth equation, which for steady state conditions and high Reynolds numbers (system sizes) reduces to the third-order cascade law~\cite{PolitanoPRE1998}. Formally
appropriate for incompressible MHD, this decomposition of the cascade is expected to be
a good approximation in weakly compressive
MHD \cite{YangEA17-mhd} and kinetic plasma
\cite{MatthaeusEA20} turbulence, 
understanding that at sub-proton scales, non-MHD effects dominate.  In homogeneous isotropic hydrodynamic
turbulence, the third-order law gives an exact relationship between
energy decay rate and the third-order structure function
\cite{Kol41c}. Originally reformulated for MHD~\cite{PolitanoPRE1998}, Hellinger et. al.~\cite{hellinger2018karman} examined the
decomposition of the von-K\'arm\'an Howarth equations for Hall-MHD.


Following standard turbulence theory~\cite{frisch1995turbulence}, the von K\'arm\'an Howarth equations may be rewritten in terms of increments $\delta \mathbf{u(r},\boldsymbol{l}) \equiv \mathbf{u(r+}\,\boldsymbol{l}) - \mathbf{u(r)}$ and  $\delta \mathbf{b(r},\boldsymbol{l}) \equiv \mathbf{b(r+}\,\boldsymbol{l}) - \mathbf{b(r)}$ where $\mathbf{r}$ is a vector in real space and $\boldsymbol{l}$ is the spatial lag, corresponding roughly to the inverse spectral wavenumber $\mathbf{k}/k^2$. The second-order structure functions are mean square values of 
these increments, 
e.g. $S_u(\boldsymbol{l}) = \langle \mathbf{
  \lvert\delta u(r},\boldsymbol{l})\rvert^2 \rangle_r$, where $\langle \dots
\rangle_r$ is an average over 
the spatial domain of
$\mathbf{r}$. Typically structure functions are averaged over lag
directions, 
giving $S_u(l) \equiv \langle\: \langle \mathbf{
  \lvert\delta u(r,}\,\boldsymbol{l})\rvert^2 \rangle_r \:\rangle_{\Omega_l},$ where $\langle \dots \rangle_{\Omega_l}$ 
  denotes averaging over solid angle~\cite{verdini2015anisotropy}. Physically, $\frac{1}{4}\,S(l) = \frac{1}{4}\,S_u(l) + \frac{1}{4}\,S_b(l)$ is the energy (flow + magnetic) inside a lag space sphere of radius $l.$

To study the cascade we employ a modified form of the
Hellinger formulation hereafter called simply the ``third-order law'' for convenience \cite{hellinger2018karman,ferrand2019exact,BandyopadhyayPRL2020}, which is averaged over direction in lag space:
\begin{equation} \label{cascadeeqn}
\frac{1}{4}\,\frac{\partial S(l)}{\partial t} + \frac{1}{4}\,\nabla_l \cdot \mathbf{Y}_l(l) + \frac{1}{8}\, \nabla_l \cdot \mathbf{H}_l(l) = \frac{1}{2}\,D(l) - \epsilon,
\end{equation}
where $\nabla_l$ is the gradient in lag space. The
MHD cascade term~\cite{PolitanoPRE1998} is $\textbf{Y}_l(l) = \hat{l}\,\left[ \hat{l} \cdot \langle\: \langle
\delta \textbf{u} \lvert \delta \textbf{u} \rvert^2 +\delta \textbf{u} \lvert
\delta \textbf{b} \rvert^2 -2\delta \textbf{b}(\delta \textbf{u} \cdot \delta
\textbf{b}) \rangle_r \:\rangle_{\Omega_l} \right]$, where $\hat{l}$ is the radial unit vector in lag space. 
Similarly, the Hall cascade term~\cite{hellinger2018karman}
is $\textbf{H}_l(l)= \hat{l}\,\left[\hat{l} \cdot \langle\:\langle 2\delta \textbf{b}(\delta \textbf{b}
\cdot \delta \textbf{j})-\delta \textbf{j} \lvert \delta \textbf{b} \rvert^2
\rangle_r\:\rangle_{\Omega_l}\right]$.
Both $\textbf{Y}_l(l)$ and $\textbf{H}_l(l)$ are mixed third-order
structure functions generalizing the hydrodynamic 
(or Yaglom) flux~\cite{monin1975statistical}.
In Eq.~\ref{cascadeeqn}, $\epsilon$ is the 
total rate of dissipation and
$D(l)$ is a lag dependent dissipation term that vanishes (by definition) outside the dissipation range. 
These are both normalized to $\omega_{ci} v_A^2$. In 
collisionless plasma simulations~\cite{matthaeus2020pathways}, the exact functional of these terms are not known. In a system with
kinematic viscosity $\nu$ and resistivity $\eta,$ 
dissipation is explicitly
$\epsilon=\nu \langle (\nabla \textbf{u}: \nabla
\textbf{u}) \rangle_r + \eta \langle \nabla \textbf{b}: \nabla \textbf{b}
\rangle_r $ and  $D(l)= \langle D(\boldsymbol{l})\rangle_{\Omega_l}$, where $D(\boldsymbol{l}) = \nu
\nabla_l^2 S_u(\boldsymbol{l}) + \eta \nabla_l^2 S_b(\boldsymbol{l})$.

\begin{figure*}[!hbt]
    \centering
    \includegraphics[scale=0.50]{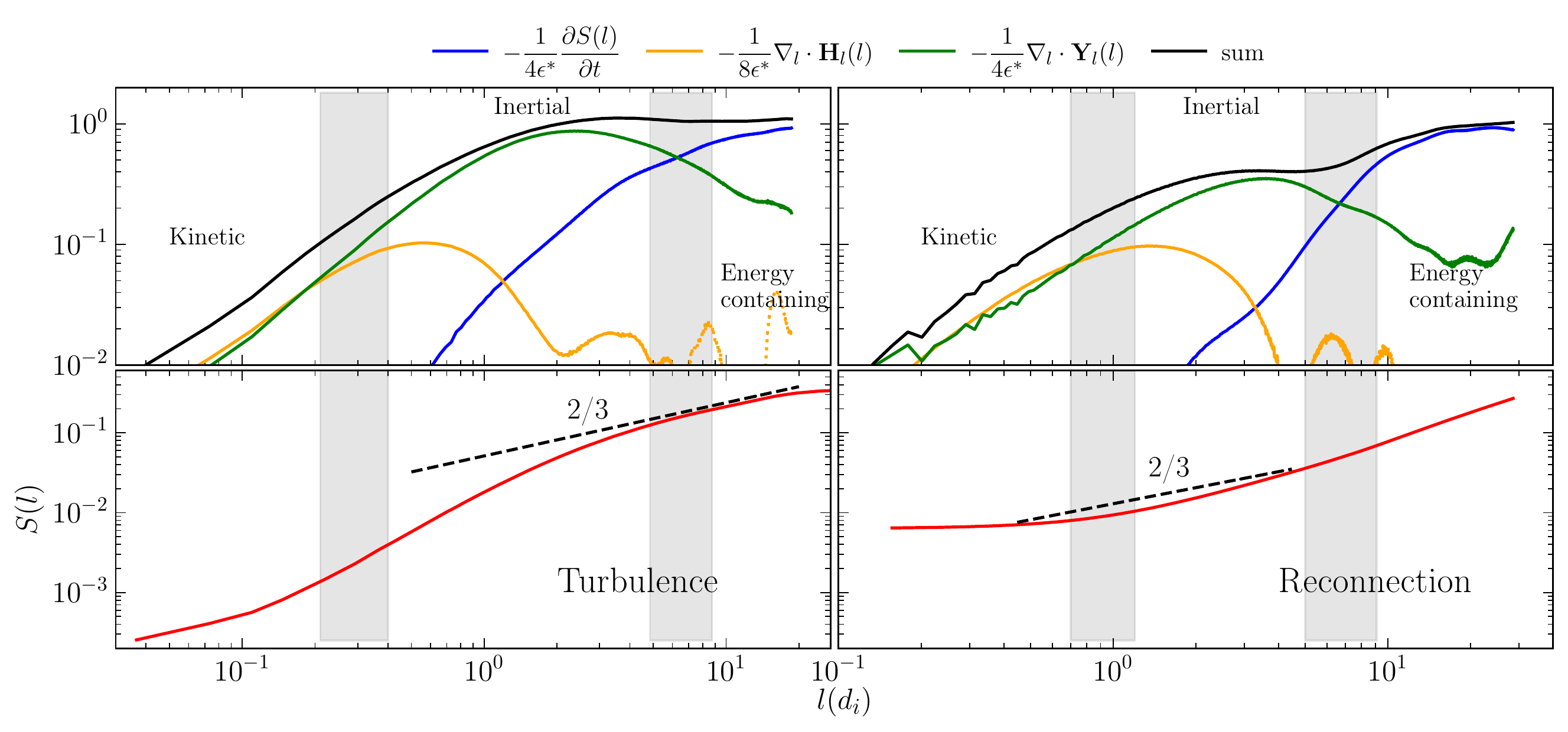}
    \caption{
    Third-order law analysis for (left) turbulence simulation when rms current is peaked ($t\omega_{ci}= 163\,$) and (right) reconnection simulation during quasi-steady phase ($t\omega_{ci} = 1045.6\,$). Top: Terms from the left hand side of Eq.~\ref{cascadeeqn} and their sum, normalized to $\epsilon^*=\partial(E_B + E_{if})/\partial t$.  Bottom: Total second-order structure function $S(l)$ with a dashed line of slope $2/3$ drawn for reference. Approximate regions of the energy containing range, inertial range, and kinetic/Hall range  are separated by light gray vertical rectangles.\\
 }
    \label{fig:third-order}
\end{figure*}

Physically, the second and third terms in Eq.~\ref{cascadeeqn} are the energy transfer rates through the surface of a lag sphere of radius $l$ due to the MHD and Hall cascades, respectively; positive (negative) is out of (into) the spherical surface. 

To study the cascade physics in our systems, we compare 
terms in Eq.~\ref{cascadeeqn} calculated from the simulations. For each time, the second and third-order structure functions are calculated as a function of lag vector ($l_x,l_y$) at each spatial grid point\cite{verdini2015anisotropy}.
The divergence is computed in lag space, and then the 1D forms are obtained using the angle averaging technique~\cite{taylor2003recovering} to give omni-directional estimates $S(l)$, $\nabla_l \cdot \mathbf{Y}_l(l),$ and $\nabla_l \cdot \mathbf{H}_l(l).$ 

For both simulations, the terms of Eq.~\ref{cascadeeqn} are time averaged over an interval ($\Delta t\omega_{ci}$) centered on the vertical red lines in Fig.~\ref{fig:sim-overview} (sim A: $\Delta t \omega_{ci} \approx 20;$ sim B: $\Delta t \omega_{ci} =22.4$ ). The average rate of change of $S(l)$ over this interval is $\partial S(l)/\partial t$ and the average rate of change of $E_B$ plus $E_{if}$ from Fig.~\ref{fig:sim-overview} gives an estimate $\epsilon^*$ for the cascade rate. $\nabla_l \cdot \mathbf{Y}_l(l),$ and $\nabla_l \cdot \mathbf{H}_l(l)$ are calculated at 5 (sim A) and 3 (sim B) evenly spaced times in the interval and averaged.

The results are shown in Fig.~\ref{fig:third-order}. The energy containing range and the inertial range are dominated by $\partial S(l)/\partial t$ and $\nabla_l \cdot \mathbf{Y}_l(l),$ respectively. In the ``kinetic/Hall'' range the Hall term $\nabla_l \cdot \mathbf{H}_l(l)$ is $\gtrsim$ the other terms. Notably, at large length scales ($l \gtrsim d_i$), the sum of the terms are constant and approximately equal to $\epsilon^*$. The constancy of the cascade rate is suggestive of the existence of an energy cascade in the system \cite{kolmogorov1991local}. $S(l)$ in Fig.~\ref{fig:third-order} exhibits an approximate slope of $2/3$ for the range of lags where the sum of the terms are constant. The results are consistent with 
a similar analysis of hybrid simulations~\cite{hellinger2018karman}. At late times when the mean square current decreases ($t \omega_{ci} \gtrsim 250\,$), the magnitude of all the contributing terms decrease (not shown), signifying a lower cascade rate. However, the region of dominance of each term persists,
resulting in a roughly constant MHD scale 
cascade rate. 

The reconnection simulation (Fig.~\ref{fig:third-order} (right)) 
exhibits many similarities to the turbulence simulation. 
In the energy containing range $\partial S(l)/\partial t$ dominates and is roughly constant, while the MHD cascade term $\nabla_l \cdot \mathbf{Y}_l(l)$ dominates 
and flattens in the inertial range; $\nabla_l \cdot \mathbf{H}_l(l)$ becomes significant approaching the kinetic range. A
feature different from the turbulence case is 
that the sum of terms exhibits two plateaus, with 
the energy containing range 
sum larger than the inertial range value. 
Also, the structure function $S(l)$ does not show a clear $2/3$ slope although it does exhibit a Kolmogorov-like $-5/3$ slope in the magnetic spectrum \cite{adhikari2020reconnection}.

\begin{figure}[t]
\includegraphics[scale=0.55]{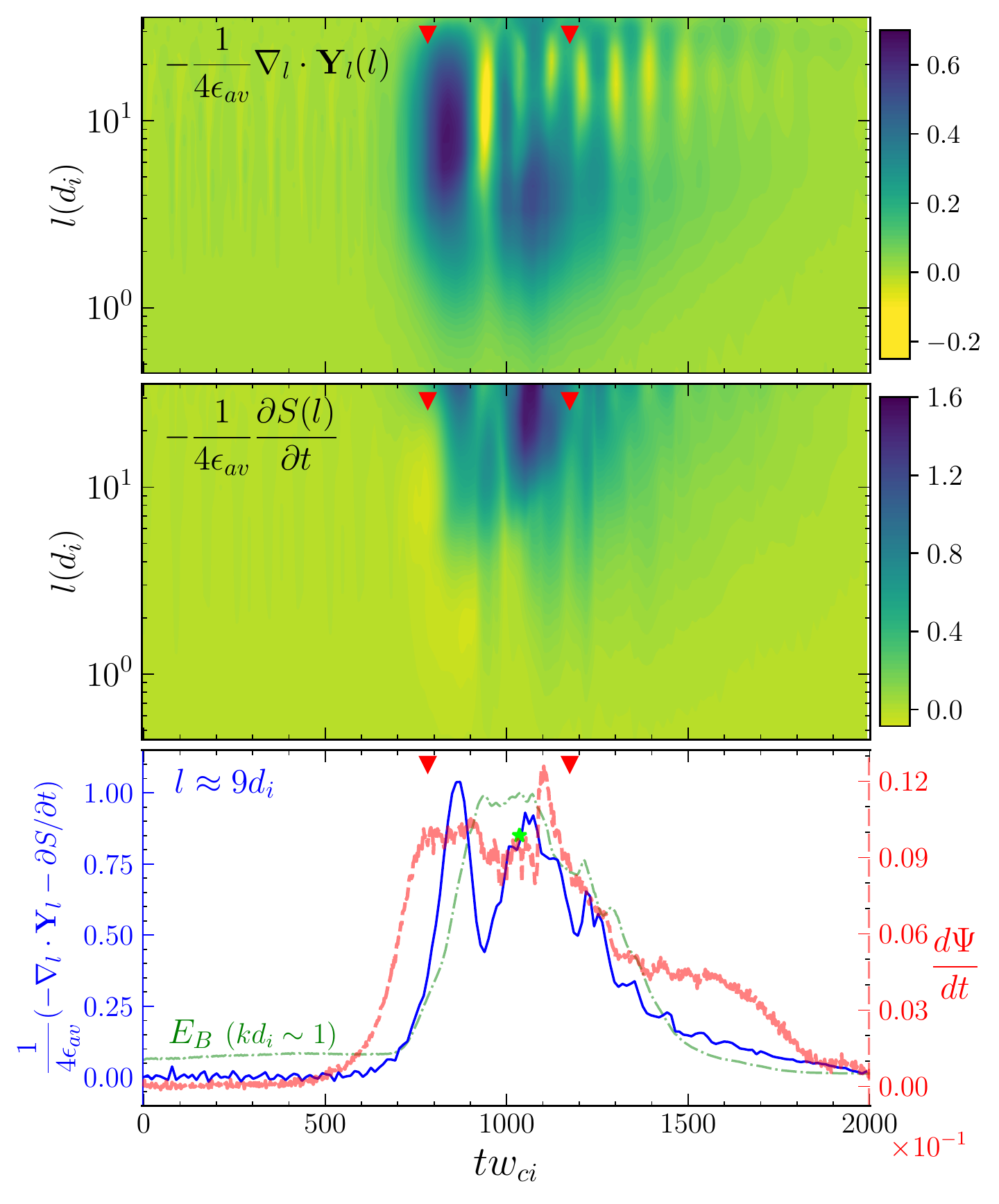}
\caption{\label{fig:time-dep-thirdorder}
Simulation B (reconnection): Time evolution of (top) $\nabla_l \cdot \mathbf{Y}_l(l)$ and (middle) $\partial S(l)/\partial t$ term, normalized to $\epsilon_{av},$ the average rate of change of $E_B$ plus $E_{if}$ during the quasi-steady period of reconnection, denoted by red triangles. (Bottom) time variation of reconnection rate (red) and $\frac{-1}{4 \epsilon_{av}}\,(\nabla_l \cdot \mathbf{Y}_l(l) + \partial S(l)/\partial t)$ at $l \approx 9\,d_i$ (blue), with the energy density $E_B(k)$ at $k\,d_i \sim 1$ (green). $E_B(k)$ plotted versus blue axis and  normalized to $(E_B(k))_{max}$. The lime star denotes $d\Psi/dt$ at $t \omega_{ci} = 1045.6$.}
\end{figure}

To strengthen the case for reconnection as a cascade process, in Fig.~\ref{fig:time-dep-thirdorder} we trace the evolution of $\nabla_l \cdot \mathbf{Y}_l(l)$ and $\partial S(l)/\partial t$ over the duration of the simulation. As reconnection initiates, the MHD cascade term develops a full inertial range for $l > d_i.$ The growth of $\partial S(l)/\partial t$ lags behind the MHD cascade term and becomes significant only after the inertial range is fully populated; the time lag between the onset of these two terms is comparable to the lag (Fig.~\ref{fig:time-dep-thirdorder} (bottom)) between the onset of reconnection (red) and the increase in spectral energy density $E_B(k)$ at $k d_i \sim 1$ (green). The MHD cascade term continues to dominate the quasi-steady phase until about $t \omega_{ci} \approx 950$ when it changes sign.

In Fig.~\ref{fig:time-dep-thirdorder} (top) for large lags, the MHD cascade term oscillates in sign
for a substantial period 
after reconnection onset. These reversals are likely related to the well-known sign-indefinite character of unaveraged
third-order correlators, along with the tendency of MHD to experience some degree of inverse transfer when close to a 2D state \cite{matthaeus1980selective,matthaeus1986turbulent,Alexakis11,coburn2015third}. Reversals of spectral flux may also be associated with non-steady $\partial S(l)/\partial t$ as suggested in Fig.~\ref{fig:time-dep-thirdorder} (middle). 
Oscillatory spectral transfer may also be triggered by a large scale Alfv\'enic  exchange between flow and magnetic energy. Notably, the oscillation period of the cascade term is roughly $90 \,t\omega_{ci}$, which is about three times the global nonlinear time of the system $\tau_{nl} = L_{box}/(2\pi\,\delta Z) \approx 33\,\omega_{ci}^{-1}.$

Following the period of quasi-steady reconnection ($t\,\omega_{ci} \gtrsim 1170$) in Fig.~\ref{fig:time-dep-thirdorder}, the MHD scale cascade rate decreases as suggested by the reduction in both $\nabla_l \cdot \mathbf{Y}_l(l)$ and $\partial S/\partial t$ in the top panels, as well as the decline of their sum for $l \approx 9\,d_i$, indicated by the blue curve in the bottom panel. The correlation of MHD cascade (blue) and reconnection rate (red), evidenced by their near simultaneous decrease, indicates a strong connection between the two processes. 

\begin{table}[t]
\fontsize{10}{12}\selectfont
\caption{\label{tab:table:cascade-rates} Dissipation and cascade rate comparisons of simulations. Left to right: correlation scale $\lambda;$ average dissipation rate $\epsilon^*$ during $\Delta t\omega_{ci}$ examined in Fig.~\ref{fig:third-order}; von K\'arm\'an cascade dissipation rate using initial $\delta Z$; von K\'arm\'an constant $C_{vk}$ (defined through $d(\delta Z^2)/dt = -C_{vk}\,\delta Z^3/\lambda \approx 2\,\epsilon^*$  ). \\
}
\begin{ruledtabular}
\begin{tabular}{ccccc}
Run & $\lambda=\frac{1}{k_{av}}[d_i] $ & $\epsilon^* [\frac{v_{A}^3}{d_i}]$ & $\frac{\delta Z^3}{\lambda} [\frac{v_{A}^3}{d_i}]$ & $C_{vk}=\frac{2\,\epsilon^*}{\frac{\delta Z^3}{\lambda}}$ \\
\hline
$A$ & $8$ & $1.7\times 10^{-4}$ & $0.011$ &  $0.031$ \\
\hline
$B$ & $14.7$ & $6.4 \times 10^{-5}$ & $0.006$ &  $0.021$ \\
\end{tabular}
\end{ruledtabular}
\end{table}


\textit{Conclusions:} Using kinetic PIC simulations, we have cross-compared the behavior of the third-order law in strong turbulence and reconnection, which allows a direct measure of the energy cascade rate. We find a significant level of structural similarities in the lag dependence of the various terms of Eq.~\ref{cascadeeqn}. Notably, both simulations exhibit an inertial range signified by a relatively 
constant and dominant MHD energy cascade term,
a signature of a turbulent system. 
Both systems have an energy containing range wherein $\partial S(l)/\partial t$ dominates and a kinetic range where the Hall term becomes important. These similarities provide evidence that the 
dynamics of reconnection proceeds through 
a cascade of energy similar to that seen in turbulence. Supporting this idea is the correlation between the reconnection rate, the sum of the MHD terms in Eq.~\ref{cascadeeqn}, and the spectral energy density at $k d_i \sim 1$ observed in Fig.~\ref{fig:time-dep-thirdorder} (bottom). 

A complimentary analysis of the reconnection and turbulence simulations is an estimate of the dissipation rate using the energy $\delta Z^2$ at the correlation scale $\lambda$; this estimate is based on the similarity decay theory of von K\'arm\'an and Howarth \cite{de1938statistical}. Table~\ref{tab:table:cascade-rates} gives the results of this analysis for the simulations and compares them to the dissipation rate $\epsilon^*$ and the MHD cascade rate $\epsilon$. The von K\'arm\'an constant $C_{vk}$ is the ratio of the actual energy dissipation to the estimated rate. While the turbulence simulation exhibits a $C_{vk}$ similar to previous PIC turbulence simulations \cite{wu2013karman}, the reconnection $C_{vk}$ is about 2/3 of the turbulence value. This reduction is almost certainly associated with the reconnection initial condition, having an energy-containing scale that is nearly at the maximum size permitted by the periodic box. This locks a fraction of the mean square magnetic potential in the largest scales, reducing the energy available to drive a direct cascade. This is the effect responsible for selective decay in $2$D MHD  \cite{MattMont80} and is analogous to Taylor relaxation in $3$D MHD.
\cite{Taylor74}. This weakens the reconnection cascade, but only fractionally.


An important point is that we have implemented the Hall-MHD third-order 
relations only for the incompressible case. While we are aware of the important 
recent developments in deriving Yaglom-like relations for compressible MHD
\cite{Andres2018energy,BanerjeeGaltier13-exact, BanerjeePRE2020}, we have not opted in the present study to implement these theories due to their general complexity, variety of forms and specialization (in some cases) to isothermal turbulence. A major issue is that the additional terms associated with compressibility do not in general fully conform to the standard structure of Yaglom laws, in which the inertial range fluxes as in Eq.{\ref{cascadeeqn}}
are in divergence form. This adds complexity to the interpretation. We suspect that it may eventually prove to be illuminating to analyze the compressive cascade separately, perhaps using scale filtering. See e.g., \cite{Aluie11,YangEA17-mhd} where it is shown that the compressible fluid and MHD  cascades are excited at large scales and proceed largely in parallel to the incompressive cascade. In any case, the presently-analysed cases are of low compressibility for which \cite{YangEA17-mhd,MatthaeusEA20} incompressible scale transfer is typically very local. 
We should also mention in passing that the cascade terms that we do include sum rather accurately to the total energy decay, scale by scale, supporting findings that MHD transfer 
is mainly local \cite{VermaEA05-mhd,AluieEyink10}. We did find some suggestion of transient back-transfer 
to long wavelengths, which is a basic property of $2$D MHD \cite{FyfeEA77b,MattMont80}, but we see no evidence 
for non-local pumping of small scales by large scales as suggested by \cite{papini2019can}.

On balance, the detailed study of reconnection from the perspective of cascade theory that we have presented here leads to what is perhaps a remarkable conclusion -- that the reconnection process leads to a standard turbulence cascade.  This does not diminish the importance of the special features of reconnection and especially its ramifications for kinetic physics. However, understanding that turbulence and reconnection are very closely related can only lead to a better understanding of each of these fundamental processes. 

We acknowledge the high-performance computing support from Cheyenne \cite{Cheyenne18} provided by NCAR's Computational and Information Systems Laboratory, sponsored by the NSF. This research also used NERSC resources, a U.S. DOE office of Science User Facility operated under Contract No. DE-AC02-05CH11231. SA, MAS, and SF acknowledge support from NASA grants NNX17AI25G, 80NSSC19K1470, 80NSSC20K0198, and NSF grant AGS-2024198. 
WHM is supported in part by the MMS Theory and Modeling team grant 
under NASA grant NNX14AC39, by NASA
Heliophysics SRT grants NNX17AB79G,
NNX17AI25G, and 80NSSC18K1648. JES is supported by Royal Society University Research Fellowship URF$\backslash$R1$\backslash$201286.



\bibliography{adhikari_cascade}


\end{document}